\documentclass{aa}  
\usepackage{natbib}

\usepackage{xcolor}

\usepackage{graphicx}
\usepackage{txfonts}
\begin{document}

\title{Observation-based modelling of the energetic storm particle event of 14 July 2012}

\titlerunning{ Modelling the ESP event of 14 July 2012}        
   \author{N. Wijsen
          \inst{1}
          \and
          A. Aran\inst{2,3}
          \and
          C. Scolini\inst{4,5}
          \and 
          D. Lario\inst{6}
          \and 
          A. Afanasiev\inst{7}
          \and 
          R. Vainio\inst{7}
          \and
          B. Sanahuja\inst{2}
          \and
          J. Pomoell\inst{8}
          \and
          S. Poedts\inst{1,9}
          }

   \institute{Department of Mathematics/Centre for Mathematical Plasma Astrophysics, KU Leuven, Belgium\\
              \email{nicolas.wijsen@kuleuven.be}
          \and
                 Departament F\'{i}sica Qu\`antica i Astrof\'{i}sica, Institut de Ci\`encies del Cosmos (ICCUB), Universitat de Barcelona (UB), Spain
            \and
                 Institut d'Estudis Espacials de Catalunya (IEEC), Barcelona, Spain
        \and
                Institute for the Study of Earth, Oceans, and Space, University of New Hampshire, NH, USA
        \and
                CPAESS, University Corporation for Atmospheric Research, Boulder, CO, USA
        \and 
                NASA, Goddard Space Flight Center, Heliophysics Science Division, MD, USA
         \and
                Department of Physics and Astronomy, University of Turku,  Finland
         \and
                 Department of Physics, University of Helsinki,  Finland
         \and
                Institute of Physics, University of Maria Curie-Sk{\l}odowska, Poland 
             }

   \date{Received 19 November 2021 / Accepted 24 December 2021}

 
  \abstract
   {}
   {
We model the energetic storm particle (ESP)  event of  14 July 2012 using  the energetic particle acceleration and transport model named `PArticle Radiation Asset Directed at Interplanetary Space Exploration' (PARADISE), together with the  solar wind and coronal mass ejection (CME) model named `EUropean Heliospheric FORcasting Information Asset' (EUHFORIA). 
  The simulation results illustrate both the capabilities and limitations of the utilised models. 
  We show that the models capture some essential structural features of the ESP event;  however, for some aspects the simulations and observations diverge.
  We describe and, to some extent, assess the sources of errors in the modelling
  chain of EUHFORIA and PARADISE and discuss how they may be mitigated in the future.

   }
   {The PARADISE model computes energetic particle distributions in the heliosphere by solving the focused transport equation in a stochastic manner. 
   This is done using a background solar wind configuration generated by the ideal magnetohydrodynamic (MHD) module of EUHFORIA. 
   The CME generating the ESP event is simulated by using the spheromak model of EUHFORIA, which approximates the CME's flux rope as a linear force-free spheroidal magnetic field. 
   In addition, a tool was developed to trace CME-driven shock waves in the EUHFORIA simulation domain. This tool is used in PARADISE to 
   (i) inject 50 keV protons continuously at the CME-driven shock and 
   (ii) include a foreshock and a sheath region, in which the energetic particle parallel mean free path, $\lambda_\parallel$, decreases towards the shock wave. 
   The value of $\lambda_\parallel$ at the shock wave is estimated from in situ observations of the ESP event.}
   {
   For energies below ${\sim} 1$ MeV, the simulation results agree well with both the upstream and downstream components of the ESP event observed by the Advanced Composition Explorer (ACE). 
   This suggests that these low-energy protons are mainly the result of interplanetary particle acceleration. In the downstream region, the sharp drop in the energetic particle intensities is reproduced at the entry into the following magnetic cloud, illustrating the importance of a magnetised CME model.
   }
  {}


   \maketitle
%

\section{Introduction}

Occasionally, particle detectors on board spacecraft measure strong intensity enhancements when an interplanetary shock driven by a coronal mass ejection (CME) crosses the spacecraft. 
Such events are referred to as energetic storm particle (ESP) events \citep{bryant62,gosling81} and are believed to be the result of continuous particle acceleration and trapping at the CME shock.
A viable acceleration mechanism is diffusive shock acceleration (DSA), which involves a self-generated turbulence process 
\citep[e.g.][]{bell78,lee83,vainio07,ng08,afanasiev15}. 
This turbulence may create a foreshock region that traps particles near the shock wave, allowing an efficient DSA process and thereby explaining the sudden particle intensity increases observed near CME-driven shocks.

Not every CME-driven shock produces detectable energetic particle intensity enhancements \citep[e.g.][]{lario03}, and it is not yet fully understood why this is the case. 
A plethora of factors are believed to contribute to the (in)efficiency of a shock as a particle accelerator, such as its  geometry and speed, the turbulence conditions near the shock wave, the presence of a suprathermal seed population, and the preconditioning produced by preceding CMEs \citep[see e.g.][for recent reviews]{desai16,reames17,guo21}. 
Many of these factors depend critically on the properties of the plasma environment through which the CME-driven shock propagates. 
In addition, the plasma conditions far from the shock wave can strongly influence the transport of the energetic particles that have escaped the shock acceleration site. 
Hence, to model a specific solar energetic particle (SEP) event, it is desirable to have an adequate description of the background plasma.  
To some extent, this can be achieved by using a global magnetohydrodynamic (MHD) model. 
An energetic particle transport model can then be used to study how the modelled CME shock  affects the transport and acceleration of particles.
Such an approach was, for example, taken by \cite{kozarev13}, \cite{schwadron15}, and \cite{young21} to model particle acceleration in the low corona using the energetic particle radiation environment module (EPREM) model \citep{schwadron10,schwadron14} together with the Alfv\'en wave solar model \citep[AWSoM;][]{vanderholst10,manchester12}.
Similarly, \citet{li21} recently used the improved particle acceleration and transport \citep[iPATH;][]{hu17} model together with the AWSoM code to model the SEP event observed on 17 May 2012. 

In this work we follow a similar approach as the aforementioned studies, but we focus on modelling the interplanetary acceleration and transport of low-energy protons (50 keV -- 2 MeV) associated with the passage by 1 au of an interplanetary CME-driven shock. This is done by using the particle transport code named `PArticle Radiation Asset Directed at Interplanetary Space Exploration'  \citep[PARADISE;][]{wijsenPHD20}, which is  coupled to the data-driven inner-heliospheric model EUHFORIA \citep{pomoell18}. EUHFORIA stands for
`EUropean Heliospheric FORcasting Information Asset'
and simulates the propagation of the solar wind and CMEs through interplanetary space by solving the ideal MHD equations using boundary conditions derived from a solar magnetogram in a semi-empirical manner. 
In \cite{wijsen21}, the EUHFORIA$+$PARADISE model was used to successfully reproduce an energetic particle event associated with a corotating interaction region (CIR), which was observed in September 2019 by both Parker Solar Probe (PSP) and the Ahead spacecraft of the Solar TErrestrial RElations Observatory (STEREO-A).  

In this work we use the EUHFORIA$+$PARADISE model to simulate the interplanetary acceleration of low-energy protons responsible for the strong ESP event observed by near-Earth spacecraft on 14 July 2012. 
The CME that generated this ESP event is simulated with the linear force-free spheromak model of EUHFORIA \citep{verbeke19,scolini19} using parameters derived from remote-sensing observations. 
To reproduce the observed intensity-time profiles at Earth, we include in the PARADISE model the effects produced by the turbulent foreshock and sheath regions formed upstream and downstream of the shock, respectively. 
Within these regions, the parallel mean free path of the energetic particles decreases towards the shock wave. 
By comparing the simulation results with in situ data, it is demonstrated that our modelling approach successfully reproduces several features of the observations. 
This comparison also highlights  a number of caveats and limitations that must be kept in mind when using MHD solar wind models together with a test-particle approach to simulate ESP and SEP events. 
We discuss each of these in detail and suggest how future developments may address them and lead to better models of SEP events with, ultimately,  predictive capabilities.

The paper is structured as follows. Section~\ref{sec:ESP-obs} contains a description of the ESP event. 
In Sect.~\ref{sec:euhforia} the EUHFORIA simulation of the solar wind and the CME associated with the ESP event is discussed. 
Section~\ref{sec:paradise} provides the details of the PARADISE simulation set-up, and in Sect.~\ref{sec:results} the results of the PARADISE simulations are discussed. Section~\ref{sec:summary} provides a summary of the present work and lists future improvements and applications of the EUHFORIA$+$PARADISE tool to consider.

\section{The ESP event on 14 July 2012}\label{sec:ESP-obs}
\begin{figure}
    \centering
    \includegraphics[width=0.49\textwidth]{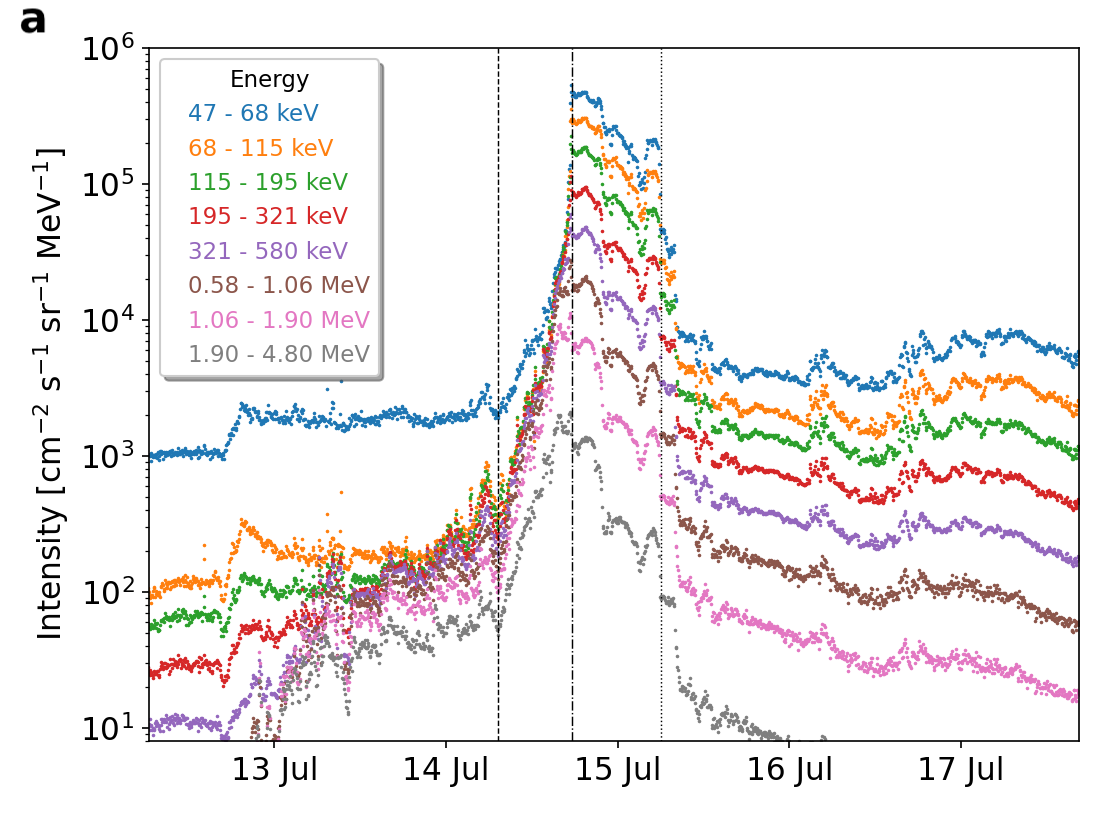}
    \includegraphics[width=0.49\textwidth]{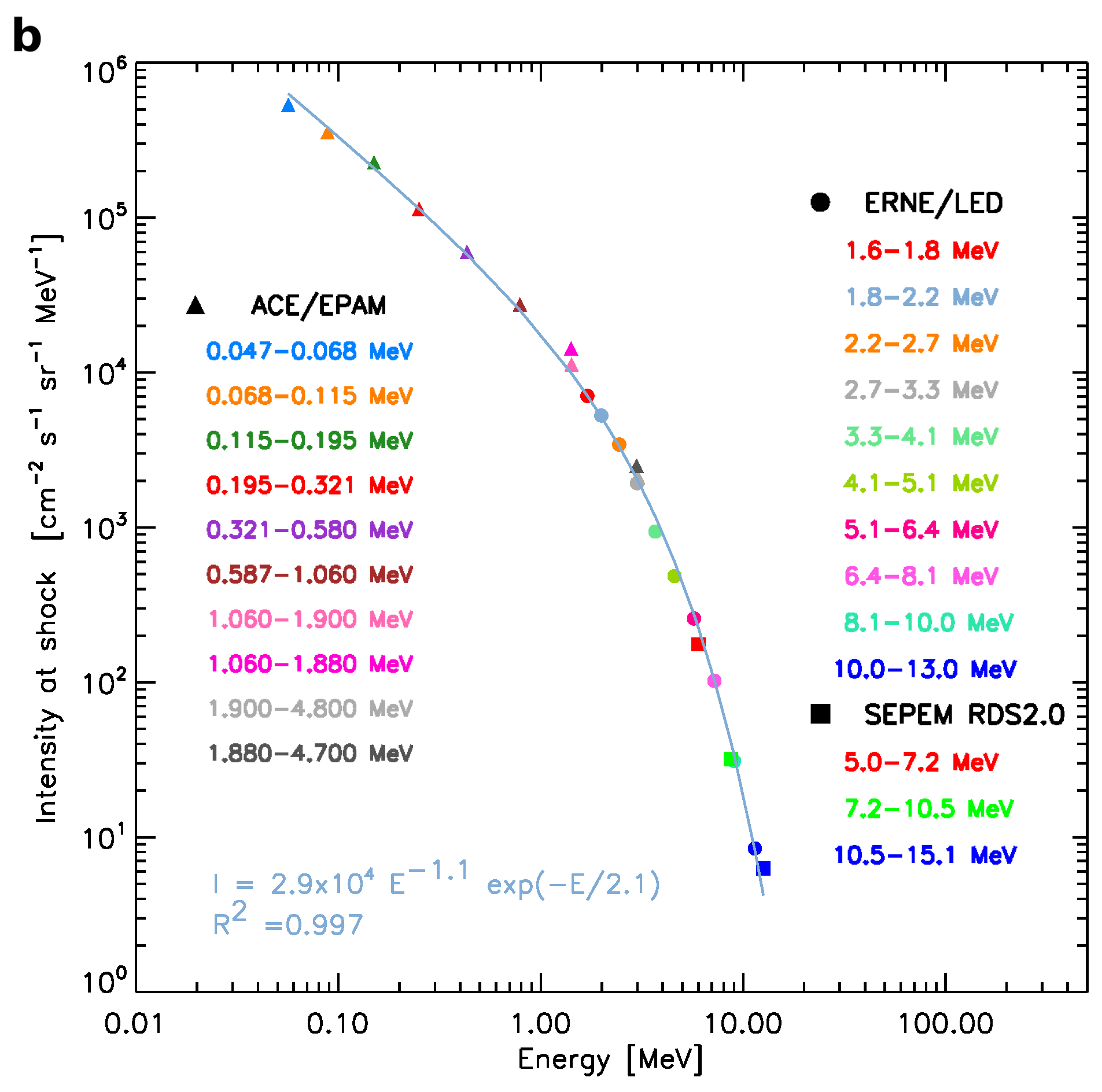}
    \caption{Ion intensity measurements of the ESP event near Earth. (a) 5-minute averages of the ion intensity-time profiles measured by ACE/EPAM for different energy channels. Far from the shock, the lowest energy channel of ACE/EPAM/LEMS120 departs from the nominal energy spectrum due to artificial counts produced by electrons bypassing the magnetic deflection system of LEMS120 \citep{marhavilas15}. The vertical lines indicate, from left to right, the onset of the ESP event, the shock arrival, and the leading edge of the flux rope. (b) Energy spectrum of the intensities observed by ACE/EPAM/LEMS120 and LEMS30, observed by ERNE/LED, and extracted from the SEPEM RDS v2.0 at the time of the shock arrival.}
    \label{fig:epam}
\end{figure}
On 14 July 2012, an ESP event was observed near Earth that lasted for approximately one day. 
As shown in Fig.~\ref{fig:epam}a, the ESP onset was characterised by abrupt and simultaneous intensity rises in all ion energy channels of the Low-Energy Magnetic Spectrometer (LEMS120) of the Electron, Proton, and Alpha Monitor \citep[EPAM;][]{gold98} on board the Advanced Composition Explorer (ACE). 
Most of the ions measured by ACE/EPAM are protons \citep{marhavilas15}.
Signatures of the ESP event were also detected by the Energetic and Relativistic Nuclei and Electron (ERNE) experiment  \citep[][]{torsti95} on board the Solar and Heliospheric Observatory (SoHO) for protons with energies up to ${\sim}30$~MeV. 
The event originated in a halo-CME that erupted from the Sun on 12 July 2012, temporally associated with a X1.4/2B flare at 15:37~UT from Active Region 11520 at S15W01. The CME arrived near Earth in the form of a flux rope with leading and trailing edges observed at $\sim$06:00~UT on 15 July and at $\sim$05:00~UT on 17 July, respectively. This flux rope was preceded by a shock observed by ACE at 17:26~UT on 14 July. 

The peak intensities of the ESP event coincided with the arrival of the CME-driven shock. 
The energy spectrum of the observed energetic particle intensities at the shock arrival time is shown in Fig.~\ref{fig:epam}b. 
To derive this energy spectrum, we included the available data from the ACE/EPAM/LEMS30 telescopes \citep{gold98} and the intensity values for $E < 15$~MeV from the SOHO/ERNE Low Energy Detector (LED) and from the Solar Energetic Particle Environment Modelling Project (SEPEM) reference data set (RDS) v2.0 \citep[sepem.eu;][]{jiggens18}. 
The energy channels used are indicated in the insets of Fig.~\ref{fig:epam}b.  
As shown in this figure, the spectrum can be fitted by a power law with an exponential rollover. 
We find that the rollover energy is located around ${\sim}2$~MeV and the exponent of the power law at lower energies is equal to $-1.1$. Moreover, as can be deduced from Fig.~\ref{fig:epam}a, the energy spectrum at low energies ($<1.90$ MeV) is relatively flat  prior to the passage of the associated interplanetary shock. 
This flattening is often observed prior to intense ESP events \citep[e.g.][]{lario18} and may result from a balance between the competing processes of DSA at the shock wave and adiabatic cooling in the upstream region \citep{prinsloo19}.

The ESP intensities measured by EPAM and ERNE showed a sudden drop on 15 July, around 06:00 UT, which corresponds to the arrival time of the leading edge of the  flux rope. During the passage of this structure, the particle intensities remained low in all energy channels.

\section{Simulating the CME with EUHFORIA}\label{sec:euhforia}

\begin{table}
\caption{Input parameters of the spheromak CME model used in the EUHFORIA simulation. }     
\label{table:euhforia}      
\centering                                      
\begin{tabular}{l l}          
\hline\hline                        
Parameter& Value \\    
\hline                                   
    Insertion time & 2012-07-12T19:24  \\
    Insertion longitude (HEEQ) & $-4^\circ$ \\
    Insertion latitude (HEEQ) & $-8^\circ$ \\
    Radius & $16.8$ R$_\odot$\\
    Density & $1.6 \times 10^{-18}$ kg m$^{-3}$ \\
    Temperature & $0.8 \times 10^6$ K \\
    Helicity & $+1$\\
    Tilt & $-135^\circ$\\
    Toroidal magnetic flux & $10^{14}$ Wb \\ 
\hline                                             
\end{tabular}
\end{table}
\begin{figure}
    \centering
    \includegraphics[width=0.49\textwidth]{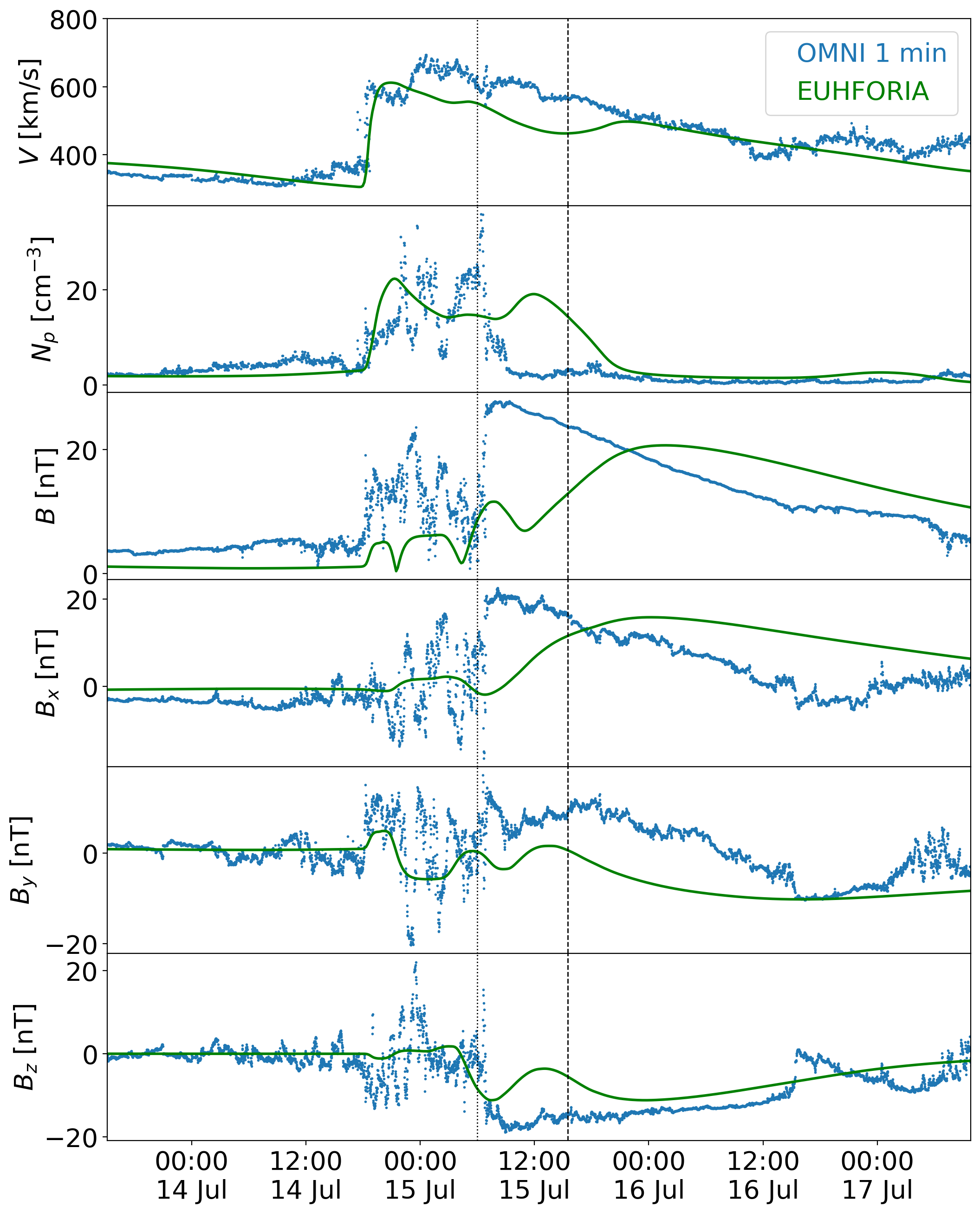}
    \caption{EUHFORIA time series at Earth (green), compared to 1 min OMNI data (blue). From top to bottom: Speed, proton number density, magnetic field strength $B$, and $B_x$, $B_y$, $B_z$ components in GSE coordinates. The vertical lines indicate the onset times of the observed (dotted line) and simulated (dashed line) flux rope. }
    \label{fig:sw}
\end{figure}
In order to model the ESP event, it is important to first have a realistic simulation of the large-scale disturbances associated with the propagating CME.
Recently, \cite{scolini19} used EUHFORIA to model the propagation of the CME through the heliosphere  for the event under study. 
This was done using the spheromak CME model, which approximates the magnetic cloud of a CME by a linear force-free spheroidal magnetic field. 
The magnetic parameters of the spheromak were estimated by \cite{scolini19} using images obtained by the Helioseismic and Magnetic Imager (HMI) instrument  and the Atmospheric Imaging Assembly (AIA) on board the Solar Dynamic Observatory \citep[SDO;][]{lemen12}. The speed and the shape of the CME were estimated from images taken by the  coronagraphs on board the STEREO-A and B spacecraft, using the graduated cylindrical shell (GCS) model \citep[][]{thernisien09, thernisien11}.
In this work, we use the same solar wind configuration and CME parameters as determined by \cite{scolini19}. 
The only difference is that the spheromak used in the present work is assumed to have a density of $1.6 \times10^{-18}$ kg m$^{-3}$ instead of $1.0 \times10^{-18}$ kg m$^{-3}$. 
This increased CME mass density provides an improved match between the arrival time of the modelled and observed CME.
The spheromak simulation parameters are summarised in Table~\ref{table:euhforia}.

Figure~\ref{fig:sw} shows the EUHFORIA simulation result at the arrival of the CME at Earth, compared to in situ solar wind and magnetic field measurements provided by the OMNI database\footnote{\url{https://omniweb.gsfc.nasa.gov/ow_min.html}} in geocentric solar ecliptic (GSE) coordinates. 
It is seen that the simulation captures the CME arrival time and the solar wind speed jump at the shock fairly well. There is also a reasonable qualitative agreement between the observed and modelled magnetic field components inside the CME. 
In Fig.~\ref{fig:sw} the arrival time of the simulated flux rope is indicated, which was determined following the methodology of \citet{scolini21}.
At the entry into the CME flux rope, the measured magnetic field exhibited a sudden increase, whereas in the simulation the arrival of the flux rope occurred ${\sim}9$ h later, showing a much more gradual increase in the magnetic field magnitude. 

\begin{figure*}
    \centering
    \includegraphics[width=\textwidth]{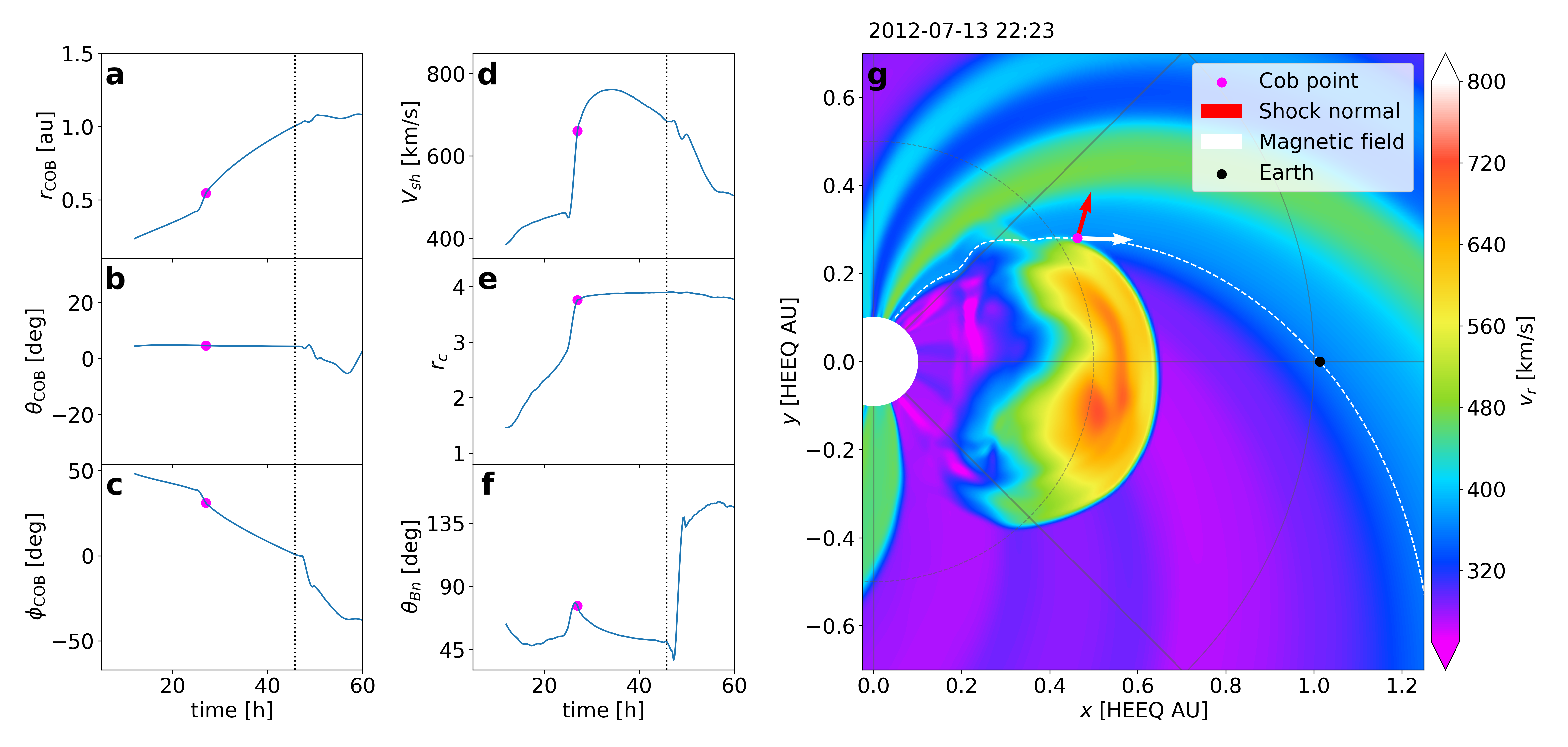}
    \caption{ Simulated shock characteristics at Earth's cobpoint. Panels~a, b, and c give, respectively, the heliospheric radial distance, latitude, and longitude of Earth's cobpoint in heliocentric Earth euatorial (HEEQ) coordinates as a function of the time elapsed  after the CME injection at 0.1~au. Panels d, e, and f give, respectively, the shock speed, the shock compression ratio, and the shock angle at Earth's cobpoint. The dashed vertical lines indicate the shock arrival time at Earth. Panel~g shows a snapshot of the solar wind radial speed at constant latitude $\theta = 4.5^\circ$ close to the time when Earth established magnetic connection with the shock front.  Earth is indicated by a black dot; it is magnetically connected to the cobpoint (magenta dot), which resides on the western flank of the CME. }
    \label{fig:shock_prop}
\end{figure*}

In  order  to  study  CME-driven  shocks  in  EUHFORIA  simulations, we developed a shock finder tool (see Appendix~\ref{app:shock_finder}). As explained in Sect.~\ref{sec:paradise}, this tool is also used to inject energetic particles continuously at the shock front and to prescribe regions of increased particle scattering in both the foreshock and the sheath region. 
Since SEPs are believed to propagate mainly along interplanetary magnetic field (IMF) lines, it is important to determine the point on the shock surface with which an observer connects magnetically. 
This point is called the cobpoint \citep[Connecting with the OBserver point; after][]{heras95}. Due to the outward propagation of the CME in space and the nominal IMF topology, the cobpoint typically moves longitudinally in clockwise direction along the shock front as seen from the north ecliptic pole. 

Figure~\ref{fig:shock_prop} shows the evolution of the cobpoint for an observer at Earth's location in the EUHFORIA simulation. 
In the simulation, this observer (hereafter Earth) connects with the shock front about $\sim$30~h after the CME insertion (i.e. on 13 July, around 22:00~UT).
However, the observed onset of the SEP event suggests that Earth had likely a direct magnetic field connection to the shock wave shortly after the CME eruption (i.e. on 12 July around 17:00~UT).
The discrepancy between the SEP event onset and the late magnetic connection in the EUHFORIA simulation can have two reasons: 
(i) the shock wave driven by the CME was in reality much wider than what is modelled and/or 
(ii) the IMF was significantly more radial than the modelled IMF. 
The latter explanation can be verified by inspecting the radial magnetic field component in Fig.~\ref{fig:sw} ($B_x$ in GSE coordinates). 
On average, EUHFORIA underestimated $B_x$ by a factor of ${\sim} 4$ in the days between the CME eruption and the shock arrival at Earth, whereas the magnitude of the  
$B_y$ and $B_z$ components were reasonably well captured by the EUHFORIA simulation. 
The strongly radial IMF might be due to a preceding high-speed stream (HSS), which crossed Earth around 11 July 2012. 
Magnetic footpoint motion at the Sun and shearing of the magnetic field in the rarefaction region may produce a so-called sub-Parker IMF \citep[e.g.][]{fisk96,fisk99,schwadron05,schwadron21}, 
for which the radial magnetic field component is significantly larger than that of a standard Parker spiral magnetic field. 
Such an IMF configuration was recently utilised by \cite{schwadron21} to explain energetic particle observations associated with a CIR. Even though EUHFORIA captures the HSS passing Earth on 11 July (green region in Fig. 3b above the Earth's location indicated by the black dot), the model does not reproduce the sub-Parker IMF. This is because EUHFORIA does not take into account the footpoint motions and the interchange reconnection that are believed to generate the sub-Parker spiral.

A parameter that strongly influences the particle acceleration efficiency at shock waves is the shock obliquity \citep[e.g.][]{ellison95},  which is determined by the angle $\theta_{B_n}$ between the upstream magnetic field and the unit normal to the shock front.
Figure~\ref{fig:shock_prop}f shows how $\theta_{B_n}$ at the cobpoint evolves from ${\sim} 85^\circ$ to ${\sim} 50^\circ$ degrees during the interplanetary propagation of the shock prior to its arrival at Earth. 
Hence, the cobpoint samples shock regions that evolve from a quasi-perpendicular to more oblique shock geometries. 
We note that a sub-Parker spiral would consistently decrease the shock angle at the cobpoint. 
For example, if we multiply the radial magnetic field component of EUHFORIA by a factor of 4, a shock angle of ${\sim} 30^\circ$ is obtained when the simulated shock crosses Earth.
In the EUHFORIA simulation, such a shock geometry is obtained in the eastern flank of the CME, to which Earth is connected while residing in the sheath. 
Furthermore, we note that just after the shock crossing, the heliospheric current sheet (HCS) crosses Earth in the simulation, explaining the ${\sim} 180^\circ$ jump in the shock angle in Fig.~\ref{fig:shock_prop}f. 

Besides the obliquity, the shock compression ratio $r_c$ also plays an important role in DSA.
Denoting the upstream and downstream solar wind velocity in a frame co-moving with the shock by, respectively,  $\vec{u}_1$ and $\vec{u}_2$, we express the shock compression ratio as $r_c := u_{1n}/u_{2n}$, where the subscript $n$ refers to the vector component along the unit shock normal.  
Figure~\ref{fig:shock_prop}e illustrates that the compression ratio $r_c$ is close to $3.9$, and remains relatively constant across the  simulated shock surface. 
Steady-state DSA predicts an energy spectrum $j \propto E^{-\Gamma}$, with spectral index $\Gamma = 0.5 (r_{\text{sc}} + 2) / (r_{\text{sc}} - 1)$ and $r_{\text{sc}}$ the scattering-centre compression ratio \citep{bell78,vainio99}. 
Under the assumption that the scattering centres are frozen into the solar wind plasma, we have that $r_{\text{sc}} = r_c$. 
Using this approximation, we find that $\Gamma =1.02$, which is slightly harder than the power law component of the observed energy spectrum shown in Fig.~\ref{fig:epam}b.

\section{PARADISE set-up}\label{sec:paradise}

To model the ESP event with PARADISE, we inject 50 keV protons continuously at the CME-driven shock wave.
Similar to \citet{prinsloo19} and \citet{wijsen21}, the injected particle intensity $j_{\rm inj}$ is scaled as
\begin{equation}\label{eq:j_inj}
    j_{\rm inj}(t,\vec{x}) \propto
       \begin{cases}
       -\nabla\cdot\vec{V}_{\rm sw}/r^2 &\text{if  $\nabla\cdot\vec{V}_{\rm sw} < 0$,}\\ 
       0 &\text{otherwise,}
       \end{cases}
\end{equation}
with $\vec{V}_{\rm sw}$ denoting  the solar wind velocity and $r$ the heliocentric radial coordinate.
This is done because $-\nabla\cdot\vec{V}_{\rm sw}$ gives a  measure of the local  compression and therefore of adiabatic heating of the background solar wind. 
A region where $-\nabla\cdot\vec{V}_{\rm sw}\gg \vec{V}_{\rm sw}/r$ may therefore contain a preheated seed population, which can more readily be injected in the DSA mechanism \citep{prinsloo19}. 
The $r^{-2}$ dependence in Eq.~\eqref{eq:j_inj} takes into account the expansion of the solar wind and hence a possible dependence for the seed population into the interplanetary medium.

It should be noted, however, that the origin of the particle seed populations of SEP events is poorly understood.  
This is also tightly coupled to the so-called injection problem of DSA \citep[e.g.][]{jokipii87}, which addresses the conundrum that particles need a sufficiently high initial energy to commence DSA. 
This is necessary to allow the particles to propagate from the downstream to the upstream shock region. 
The particles' injection energy is expected to depend strongly on the shock obliquity and the particle diffusion conditions \citep[see e.g.][]{giacalone99,zank06,neergaardParker12}. 
In particular, the injection energy may be significantly higher for a quasi-perpendicular shock, unless the particles are subjected to an efficient cross-field diffusion process \citep[e.g.][]{zank06,neergaardParker14}.
By injecting 50~keV protons in the PARADISE simulation, we do not address any effect of the energy spectrum of the seed population on the modelled particle acceleration. 
However, since the injected protons get accelerated in the simulation, we can infer that 50~keV is above the injection energy of the modelled shock under the assumed diffusion conditions.

PARADISE solves the focused transport equation \citep[FTE; e.g.][]{roelof69,isenberg97} including a spatial diffusion process perpendicular to the IMF and a pitch-angle diffusion process. 
The  pitch-angle diffusion process results from the interaction between the energetic particles and Alfv\'enic slab-like turbulence. 
PARADISE uses the results of quasi-linear theory \citep[QLT;][]{jokipii66} to prescribe the following pitch-angle diffusion coefficient \citep[see][]{agueda13,wijsen19a}:
\begin{equation}
    {D_{\mu\mu} = D_0\left(\frac{|\mu|}{1 + |\mu|} + \epsilon\right)\left(1-\mu^2\right)},
\end{equation}
where $\mu$ denotes the cosine of the pitch-angle, $\epsilon = 0.048$ is a parameter bridging the resonance gap at $\mu = 0$ \citep[e.g.][]{klimas71}, and $D_0$ is a scaling factor that depends on the particle's rigidity.
In this work, $D_0$ is determined by prescribing the particles' parallel mean free path $\lambda_\parallel$, which relates to $D_{\mu\mu}$ through \citep{hasselmann70}
\begin{equation}
    \lambda_\parallel = \frac{3\varv}{8}\int_{-1}^{1}\frac{\left( 1 - \mu^2\right)^2}{D_{\mu\mu}}d\mu,
\end{equation}
with $\varv$ denoting the particle speed.
During ESP events, the level of slab turbulence is typically enhanced in front of the shock wave due to the amplification of Alfv\'en waves by the energetic particles that are undergoing acceleration \citep{lee83}. 
Similarly, the levels of turbulence are also expected to be enhanced in the sheath as compared to the ambient solar wind \citep[e.g.][]{masias-meza16}.
As a result, both the foreshock and the sheath may be characterised by a small $\lambda_\parallel$. 
Since this plays a crucial role in the production of ESP events, it needs to be included in the PARADISE simulations. 
This is achieved by prescribing a parallel mean free path of the form
\begin{align}
\lambda_\parallel^{\text{IP}} &= (0.1~\text{au}) \left(\frac{R}{R^{\rm ref}}\right)^{\beta} \label{eq:mfp_IP},\\
 \lambda_\parallel^{\text{foreshock}} &= \min \left(f(x)\frac{\cos^2\theta_{B_n}^{\rm ref}}{\cos^2\theta_{B_n}} \left(\frac{R}{R^{\rm ref}}\right)^{\beta},\lambda_\parallel^{\text{IP}}\right)\label{eq:mfp_fs},\\ 
  \lambda_\parallel^{\text{sheath}} &= \min\left(f(d) \left(\frac{R}{R^{\rm ref}}\right)^{\beta},\lambda_\parallel^{\text{IP}}\right) \label{eq:mfp_ss},\\ 
 f(s) & =  (as+b) e^{s/\Delta_{\rm feb}} \label{eq:f(s)},
\end{align}
where $\lambda_\parallel^{\text{IP}}$ is the interplanetary parallel mean free path, $R$ is the particle rigidity, $d$ is the distance to the shock, $x$ is the distance to the shock measured along the IMF, $\Delta_{\rm feb}$ represents a free escape boundary, and $a$, $b$, and $\beta$ are free parameters which are specified below. 
The reference rigidity is chosen as $R^{\rm ref} = 12.88$~MV, corresponding to $88$~keV protons. The assumed rigidity dependence is based on QLT and a Kolmogorov turbulence spectrum, that is, we choose $\beta =1/3$ unless specified otherwise.  The reference shock angle is chosen as $\theta_{B_n}^{\rm ref} = 49^\circ$,  equal to the modelled shock angle when the CME arrives at Earth. 
The free-escape boundary is assumed to be $\Delta_{\rm feb} = 0.17$~au. This value was obtained by multiplying the modelled shock speed at Earth (${\sim} 700$ km/s) by the duration of the observed ESP onset phase (${\sim} 10$~h). 
The chosen value for $\Delta_{\rm feb}$ is also close to the presumed width of the sheath. 
This width  can be estimated by taking into account that the  arrival times of the shock and the leading edge of the flux rope are ${\sim} 12$~h apart. 
Multiplying this by the shock speed gives a width of the sheath of  ${\sim} 0.2$~au.

The linear dependence of $f(s)$ close to the shock is based on Eq.~(22) of \citet{vainio14}. In that work, the authors construct a semi-analytical model for ESP events, by using the analytical model of DSA derived by \citet{bell78} as a starting point and re-calibrating the theoretical functional forms using numerical simulations of self-generated waves.
Furthermore, we included a dependence on $\cos\theta_{B_n}$ in $\lambda_\parallel^{\text{foreshock}}$ because quasi-parallel shocks are expected to have a more turbulent foreshock. 
This is because the quasi-parallel geometry allows particles to escape more easily upstream of the shock, where they subsequently induce beam-driven instabilities that excite various plasma waves that are expected to decrease the parallel mean free path of the particles \citep[see e.g.][ and reference therein]{balogh13}.


When using an MHD model such as EUHFORIA to simulate the evolution of a CME in the inner heliosphere, the shock wave generated in front of the CME will typically be thicker than a real interplanetary shock wave, due to the limited resolution of the numerical grid. 
As explained in Appendix~\ref{app:particle_acceleration}, prescribing a small $\lambda_\parallel$ across the simulated shock would lead to compressional acceleration instead of first-order Fermi acceleration.
The former happens because the diffusion length scale is smaller than the shock width \citep[e.g.][]{schwadron20, wijsenPHD20}. 
To avoid this and hence to allow particles to cross the simulated shock multiple times, we allow the particles to cross the shock wave in a scatter-free manner, 
that is, the particles propagate across the modelled shock structure while conserving their magnetic moment. 
Hence, unless a particle is mirrored by the magnetic compression, it will cross the entire shock wave structure and sample the full compression ratio. 
Since the particles still need a finite time to interact with the modelled shock wave, the acceleration efficiency of the modelled DSA is expected to be slightly reduced as compared to the case where the shock is a sharp discontinuity.

\begin{figure}
    \centering
    \includegraphics[width=0.48\textwidth]{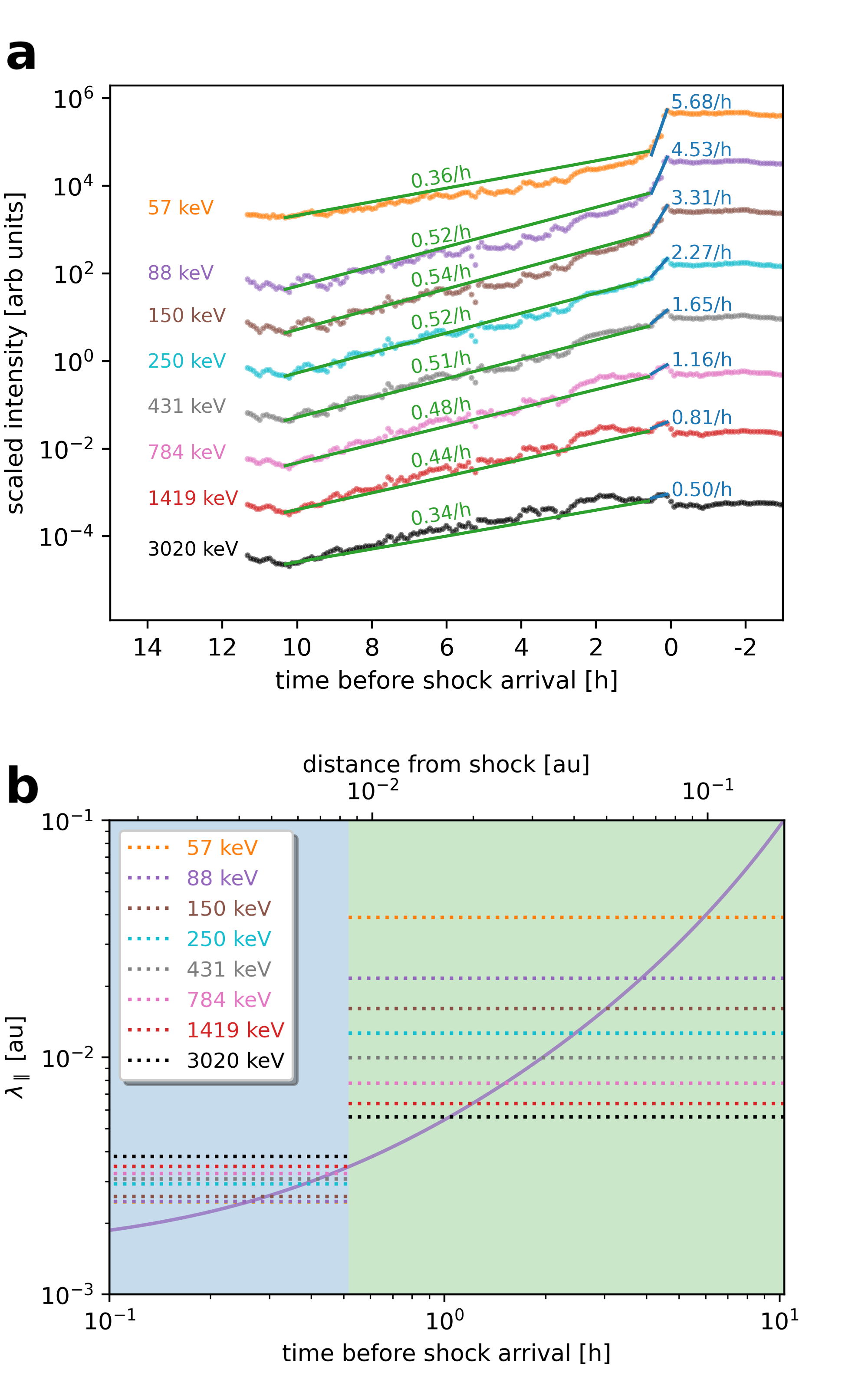}
    \caption{Parallel mean free path in the foreshock derived from in situ ion intensity measurements. 
    In panel (a) dots give the intensity-time profiles observed by EPAM for different energy channels, shifted downwards to avoid overlap. 
    The straight lines give a time-averaged increase in the time intervals $T_1 = [0{\text{h}}, 0.5{\text{h}}]$ (blue) and $T_2 = [0.5{\text{h}},10.3{\text{h}}]$ (green), and the values above each line give the slope. 
    In panel (b) the dotted lines give the $\lambda^{\rm DSA}_\parallel$ derived from the slopes presented in (a) for $T_1$ (blue background) and $T_2$ (green background). The solid line gives the $f(s)$ (see Eq.~\eqref{eq:f(s)}) used in the PARADISE simulations.   }
    \label{fig:mfp}
\end{figure}

The parameter $b$ in Eq.~\eqref{eq:f(s)} gives the parallel mean free path for $88$~keV protons at the shock ($x=0$), and can be estimated from the measured particle intensities. 
This is done by approximating the observed intensity-time profiles by a function of the form $f = \alpha e^{ t/t^*}$, with $t^*$ being the $e-$folding timescale. 
This procedure is illustrated in Fig.~\ref{fig:mfp}a for two different time intervals prior to the shock arrival (i.e. $ T_1=  [0{\text{h}}, 0.5{\text{h}}]$ and  $ T_2 = [0.5{\text{h}},10.3{\text{h}}]$).
We note that the $57$~keV energy channel has a high background level, likely affecting the intensity-time  profiles in the first two time-intervals. 
Figure~\ref{fig:mfp}a  also shows that the intensity time-profiles remain approximately constant after the shock passage, which is in agreement with DSA \citep[e.g.][]{bell78,drury81}.
Assuming constant spatial diffusion coefficients $D_n^{\rm DSA}$ in the two time intervals, steady state DSA can be used  to express $D_n^{\rm DSA}$ as
\citep[e.g.][]{vannes84,beeck89,giacalone12c}
\begin{equation}\label{eq:Dn}
     D_n^{\rm DSA}(t,E) =  -u_{n1} t^* V_{\rm sh},
\end{equation}
where $V_{\rm sh}$ is the shock speed. 
The corresponding parallel mean free path $\lambda^{\rm DSA}_\parallel$ can then be calculated as
\begin{equation}
    \lambda_{\parallel}^{\rm DSA}(t,E) = \frac{3D_n^{\rm DSA}}{\varv\cos^2 \theta_{B_n}}.
\end{equation}
The values for $\lambda^{\rm DSA}_\parallel$ obtained for the different energy channels are shown as dotted lines in Fig.~\ref{fig:mfp}b. 
To obtain these values, the EUHFORIA simulation was used to prescribe $\theta_{B_n}=49^\circ$, $V_{\rm sh} = 683$~km\,s$^{-1}$, and $u_{n1}=-401$~km\,s$^{-1}$. 

The values obtained for $\lambda_\parallel^{\rm DSA}$ should be treated with care as steady-state DSA does not take the time history of the (fore)shock or the effect of magnetic focusing due to the diverging IMF into account.
The latter process may especially affect the intensity-time profiles in the time interval $T_2$. For this reason, we determine the parameters $a$ and $b$ solely based on the first interval $T_1$, where the diffusive approximation is assumed to be more valid.
We chose the parameters $a$ and $b$ such that
\begin{equation}
 \begin{aligned}
     f(t = 0.25~{\text{h}}) &= \lambda_{\parallel}^{\text{DSA}}(T_1,88~\text{kev}) = 2.4\times 10^{-3}~\text{au},\\
    f(s = \Delta_{\rm feb}) &= 0.1~\text{au},
 \end{aligned}
\end{equation}
where $t= 0.25~{\text{h}}$ is the midpoint of $T_1$. This gives $b = 1.1\times10^{-3}$~au and $a =2.1\times 10^{-2}$. 
The resulting mean free path in the foreshock for $88~$keV protons and $\theta_{B_n} = 49^{\circ}$ is shown as a solid line in Fig.~\ref{fig:mfp}b
(see also Fig.~\ref{fig:mfp_on_sw}).
The parameters $\Delta_{\text{feb}}$, $a$, and $b$ are kept constant during the entirety of the simulation. 
Hence, Eqs.~\eqref{eq:mfp_IP}~--~\eqref{eq:f(s)}  only contain an implicit time dependence due to the propagation and expansion of the shock wave and due to the dependence of $\lambda^{\rm foreshock}_\parallel$ on the shock obliquity. 
In reality, $\lambda_\parallel$ is expected to be time dependent due to the coupling between the accelerated energetic particle distributions and the turbulence \citep[e.g.][]{lee83}. Capturing the wave-particle dynamics is, however, outside the scope of the current study, but will be the subject of future improvements to our model.

Our simulations include a weak cross-field diffusion process, which is characterised by a constant perpendicular mean free path $\lambda_\perp = 10^{-4}$~au. Hence, we have that $\lambda_\parallel / \lambda_\perp > 1$ everywhere in the simulation, meaning that parallel particle transport will dominate. We leave it for future work to study the effect of different cross-field diffusion conditions. 

Finally, we note that the EUHFORIA snapshots are updated every 15 minutes in PARADISE.
Apart from the MHD plasma variables, each snapshot contains the location of the shock wave and $\lambda_\parallel$, computed according to Eqs.~\eqref{eq:mfp_IP} -- \eqref{eq:f(s)}. 
A linear interpolation in time is performed to obtain the plasma variables and $\lambda_\parallel$ at times between two consecutive snapshots. 

\section{Results}\label{sec:results}

\subsection{Intensities at different longitudes}
\begin{figure}
    \centering
    \includegraphics[width=0.49\textwidth]{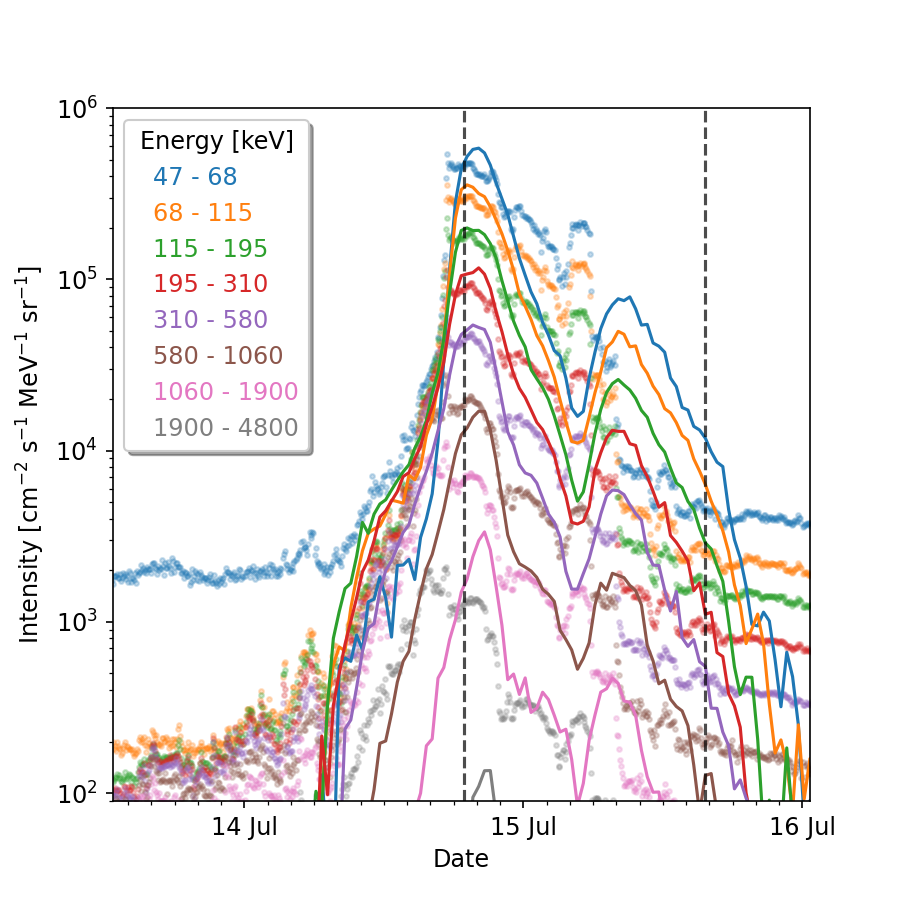}
    \caption{Simulated (solid lines) and observed (dots) omnidirectional intensity-time profiles at Earth. The dots in the background give the ACE/EPAM/LEMS120 data (see also Fig.~\ref{fig:epam}a). The dashed vertical lines indicate the arrival times of the shock (left) and the flux rope (right) in the EUHFORIA simulation.}
    \label{fig:I_earth_sim}
\end{figure}

Figure~\ref{fig:I_earth_sim} shows the simulated omnidirectional intensity-time profiles at Earth. 
The particle distribution is scaled such that the simulated peak intensity matches the observed EPAM peak intensity in the 68.1 -- 115 keV channel.  The figure shows that the PARADISE simulation successfully reproduces several features seen in the EPAM data, especially for energies below 1 MeV.
For these energies, the simulation reproduces: 
(1) the onset time of the ESP event, 
(2) the intensity increase prior to the shock arrival, 
(3) the energy spectrum at the shock wave, 
(4) the intensity drop near the onset of the magnetic cloud, 
and (5) a secondary intensity peak just before the onset of the magnetic cloud. 
Both in the simulation and in the data, the secondary peak occurs just after the three magnetic field components become negative and subsequently attain a local minimum.
This structure is more smeared-out in the EUHFORIA simulation as compared to the data, explaining why the secondary intensity peak in the PARADISE simulation is lower in intensity but spans a longer time interval than in the EPAM data. 

\begin{figure*}
    \centering
    \begin{tabular}{c c}
     \includegraphics[width=0.4\textwidth]{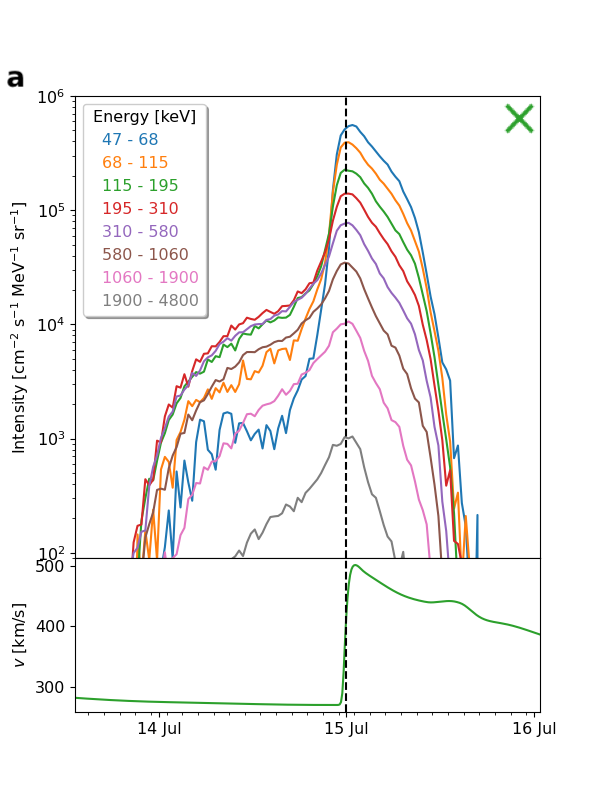}  &  \includegraphics[width=0.4\textwidth]{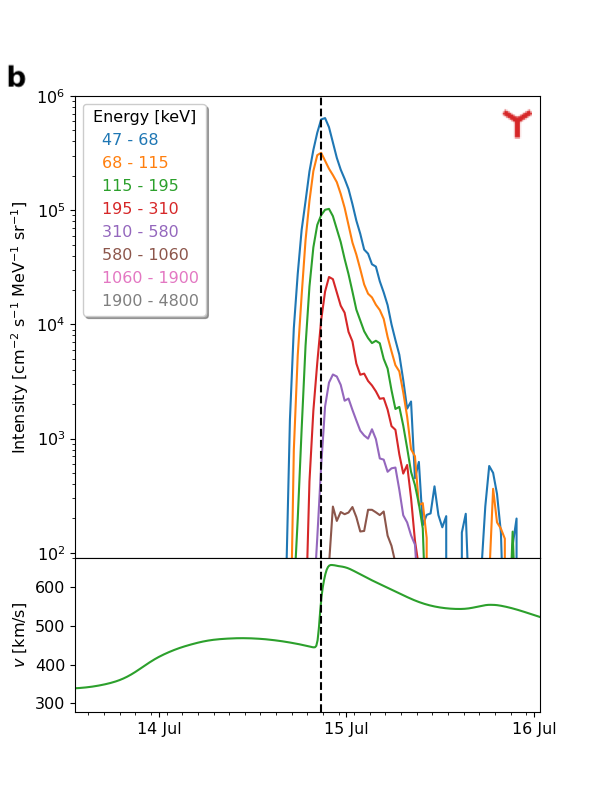}  \\
         \includegraphics[width=0.45\textwidth]{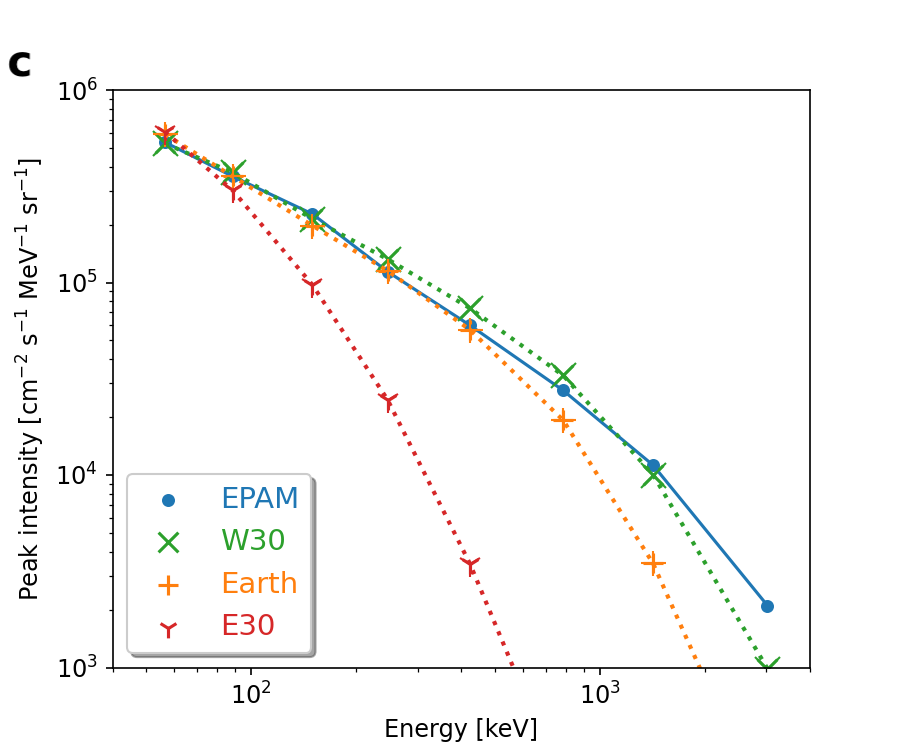} &
          \includegraphics[width=0.48\textwidth]{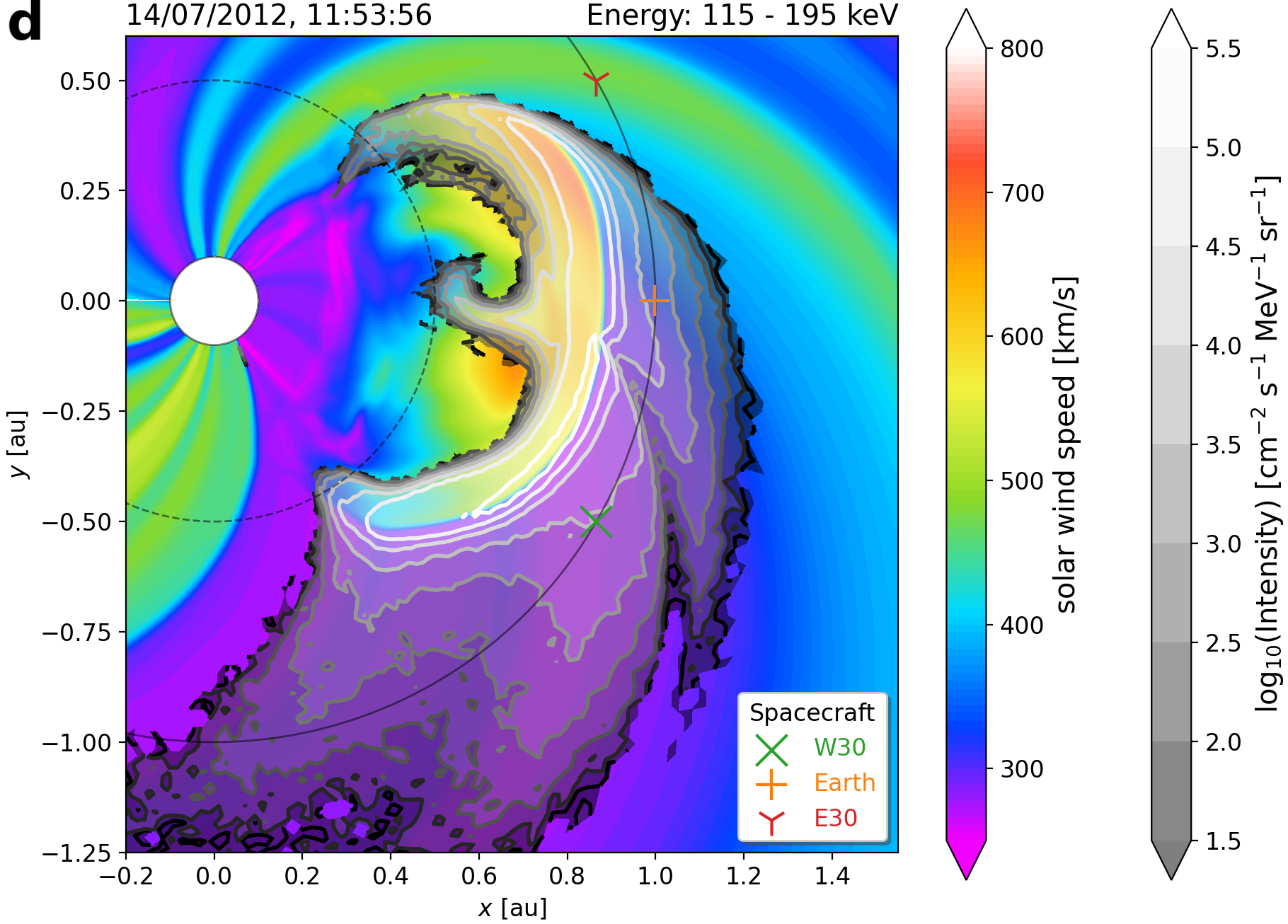}
    \end{tabular}
    \caption{Simulation results for virtual observers at different longitudes. Panels \textbf{(a)} and \textbf{(b)} give the omnidirectional intensity-time profiles (upper panel) together with solar wind speed (lower panel) for the W30 and E30 observers, respectively. Panel \textbf{(c)} gives the peak intensities versus energy for different observers.  Panel \textbf{(d)} is a snapshot of the omnidirectional particle intensity plotted in grey shades on top of the solar wind speed at constant HEEQ latitude $\theta = 4.5^\circ$.   }
    \label{fig:I_lons}
\end{figure*}

The simulation captures only the ESP event, and not the preceding SEP component that started on 12 July at around 17:00~UT. 
This is due to the late (by ${\sim} 30$ h) magnetic connection of Earth to the simulated shock front (see Sect.~\ref{sec:euhforia}). 
Moreover, the peak intensities of the high-energy channels are reached after the shock arrival at Earth. 
This is because after the shock passage, the cobpoint of Earth moves to the eastern flank of the CME, where the acceleration is more efficient due to a more parallel shock geometry. 
To illustrate this, we show in Fig.~\ref{fig:I_lons} the intensity-time profiles of two virtual observers located at the same radial distance and latitude as Earth but shifted in longitude, that is, one virtual spacecraft located $30^\circ$ east of Earth, which we  refer to as the E30 observer, and another  
virtual spacecraft located $30^\circ$ west of Earth, referred to as the W30 observer. 
These observers are indicated by the green and red symbols in Fig.~\ref{fig:I_lons}d, which shows the omnidirectional particle intensities together with the solar wind speed in a slice of constant latitude, approximately 6h before the shock arrival at Earth. The figure illustrates that the particle peak intensities are obtained at the CME shock wave. Moreover, the highest particle intensities are obtained on the eastern CME wing, where the shock is quasi-parallel. 

The W30 observer (green symbol) has an early connection to the shock wave, and hence starts observing energetic particles already  on 14 July, around 00:00 UT.
This is still more than a day later than the onset of the SEP event observed at Earth. 
This is because we do not include particle acceleration and transport happening in the corona and we only start injecting 50~keV protons at the time the CME shock is formed in the EUHFORIA domain. 
Figure~\ref{fig:I_lons}a shows that during the simulated event onset, the energy spectrum is increasing up to 115~keV, flat from 115~keV to 580 keV, and decreasing for higher energies. 
This can be attributed to the non-trivial interplay between the acceleration efficiency at the shock, the particle escape efficiency from the foreshock, the adiabatic deceleration in the solar wind, and the time a particle needs to travel from the shock to the spacecraft. All four processes are energy dependent. 

In contrast to the simulated intensity profiles at Earth, the particle intensities peak in all energy channels at the time of the shock arrival at  the W30 observer. Figure~\ref{fig:I_lons}c also reveals that the energy spectrum of the W30 observer is slightly harder than at Earth, despite the fact that the shock encountered by the W30 observer is slightly weaker ($r_c = 3.8$) than at Earth ($r_c = 3.9$). 
This can be attributed to the more parallel shock geometry of the eastern CME wing. The inverse proportionality of $\lambda_\parallel^{\rm foreshock}$ on $\cos\theta_{B_n}$ (see Eq.~\eqref{eq:mfp_fs}) allows particles to be more easily trapped at a quasi-parallel shock, increasing the probability on multiple shock crossings.

Due to its position, the E30 observer (red symbol) connects around 05:00 UT on 14 July to the CME's western flank, which is characterised by a quasi-perpendicular shock geometry. 
Along this flank, the acceleration is less efficient partly due to the small foreshock, which is because of the $\cos\theta_{B_n}$ dependence in Eq.~\eqref{eq:mfp_fs}. 
In addition, the particles residing on the field lines crossing the western flank of the CME only had limited time to interact with the shock wave and hence get accelerated. 
As for the observer at Earth, the high-energy channels peak after the shock passage, which is because the cobpoint of the E30 observer moves towards the shock nose.  

\subsection{Intensities at different radial distances}

\begin{figure}
    \centering
    \includegraphics[width=0.49\textwidth]{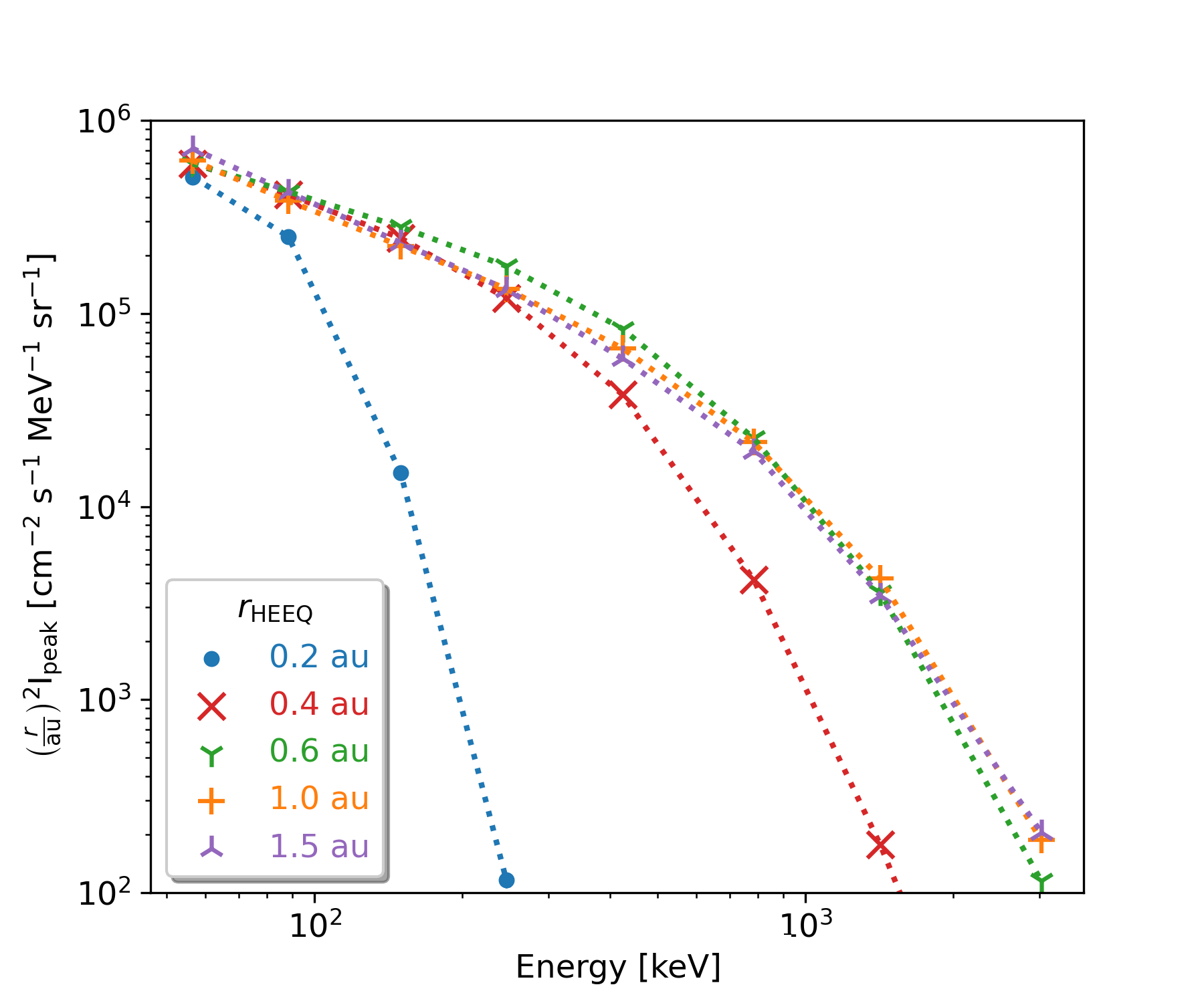}
    \caption{Peak intensities scaled by $r^2$ versus energy for virtual observers located along the Sun-Earth line.}
    \label{fig:ES_radial}
\end{figure}

Figure~\ref{fig:ES_radial} shows the energy spectra measured at the peak intensities for different observers located along the Sun-Earth line, hence illustrating the time-evolution of the particle energy spectrum at a point close to the shock nose. 
The peak intensities have been scaled by a  $r^2$ factor, to cancel out the effect of the solar wind expansion, which is incorporated in our injection distribution through the $r^{-2}$ dependence in Eq.~\eqref{eq:j_inj}.
It is seen that the energy spectrum hardens with radial distance up to ${\sim} 0.5$~au. 
After that, the shape of the energy spectrum remains approximately constant and can be characterised by a power-law with an exponential rollover at $\sim 700$~keV.
Although not shown here, the energy spectrum shows a similar evolution for observers located along the Sun-W30 line. 
In this case, the exponential rollover is located around $1$~MeV (see also Fig.~\ref{fig:I_lons}c).
In future studies, we plan to compare the modelled radial behaviour of the energy spectrum with observations from PSP \citep{fox16} and Solar Orbiter \citep[SolO;][]{muller20}. 

\subsection{Intensities for different rigidity dependences}

\begin{figure}
    \centering
    \includegraphics[width=0.49\textwidth]{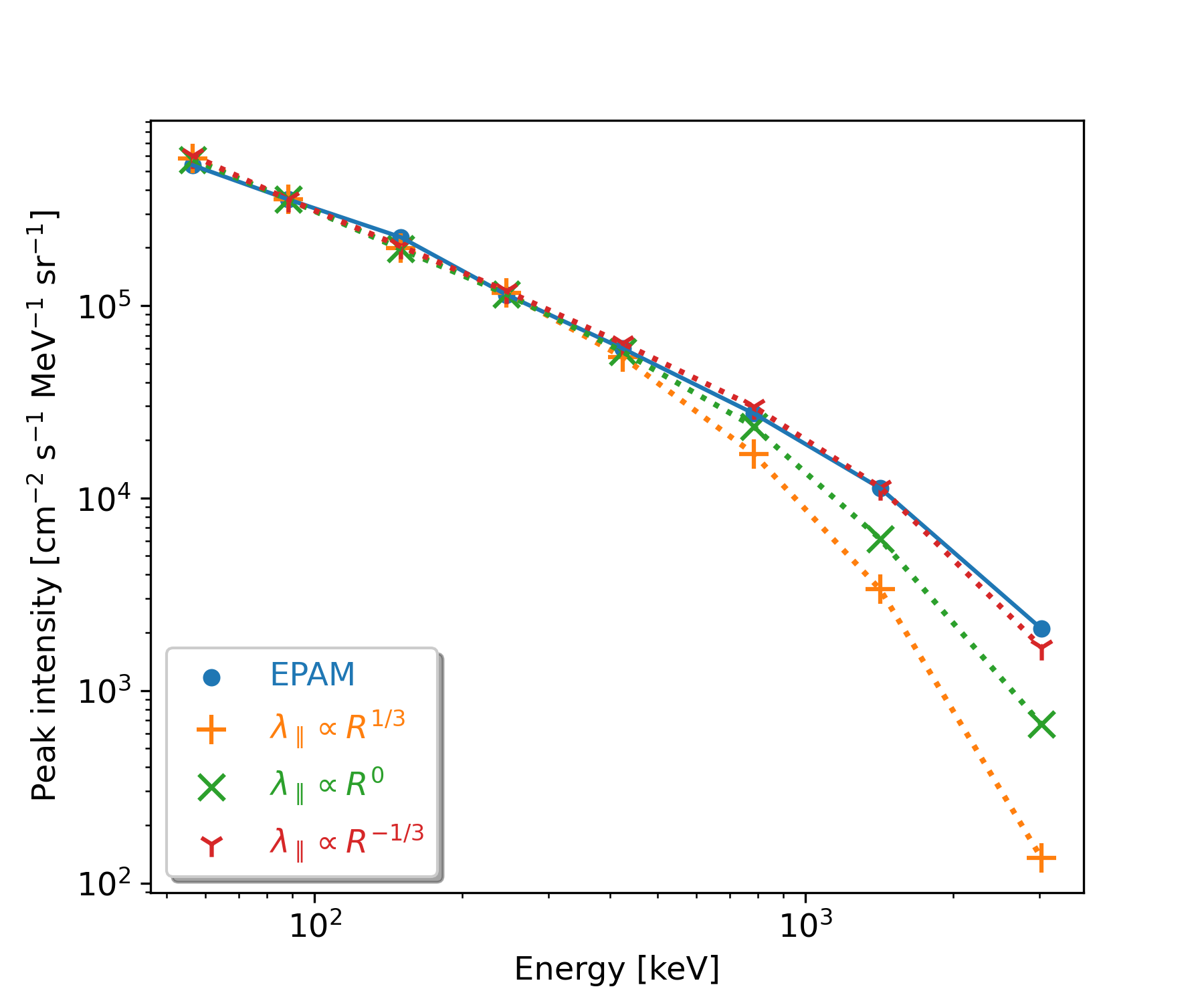}
    \caption{Peak intensity energy spectra obtained from different $\lambda_\parallel$ power-law dependences on the ion rigidity. 
    The orange, green, and red symbols give the results of PARADISE simulations using the indicated rigidity dependence. Blue symbols give the peak intensities measured by EPAM.}
    \label{fig:ES_rigid}
\end{figure}

In the simulation results presented above, the parallel mean free path followed the QLT  rigidity dependence (i.e. $\beta = 1/3$ in Eqs.~\eqref{eq:mfp_IP} -- \eqref{eq:f(s)}). 
This means that the parallel mean free path increases with the particle energy. 
However, in the steady-state DSA theory derived by \citet{bell78}, the parallel mean free path decreases with increasing energy for a shock wave with $r_c \sim 4$.
\citet{beeck89} analysed five in situ observed foreshock regions and found both positive and negative rigidity dependences.
 This result was obtained using a method similar to the one presented in Sect.~\ref{sec:paradise} (see Eq.~\eqref{eq:Dn}).
Moreover, by calculating the scattering mean free path directly from
the observed magnetic field fluctuations, \citet{beeck89} obtained comparable rigidity dependences and mean free path values when assuming that the particle scattering resulted from a combination of Alfv\'en waves and magnetosonic waves. 
In contrast, when only considering the resonance scattering due to Alfv\'enic fluctuations, the authors derived smaller mean free paths and positive rigidity dependences.
Such a positive dependence was also found by \citet{afanasiev15}  using the  SOLar Particle Acceleration in
Coronal Shocks (SOLPACS) model,
which takes into account the resonant interactions between energetic particles and Alfvén waves in the foreshock, under the assumption of QLT.
The main difference of this modelling to the theory of \citet{bell78} (in addition to time dependence) is that the wave-particle resonance condition is taken into account in full, accounting for the effect of the pitch angle that is omitted in the theory. 
This leads to low-energy protons being scattered also by waves generated by high-energy protons especially further out from the shock.

To better understand the effect of the rigidity dependence on the results presented in this work, we performed simulations with $\beta$ equal to $1/3$, $0$, and $-1/3$.  
Figure~\ref{fig:ES_rigid} shows the peak intensity energy spectra obtained by a spacecraft located at Earth for these three simulations. 
It can be seen that decreasing $\beta$ produces a harder energy spectrum. 
This is because the high-energy particles become more efficiently trapped near the shock and hence they have a larger probability of multiple shock encounters. 
The acceleration efficiency of the high-energy particles is thus increased.
The $\beta = -1/3$ case gives the best correspondence with the EPAM data. However, this contrasts with the positive energy dependence derived from the intensity observations at 1 au in time interval $T_1$, using steady state DSA (see Fig.~\ref{fig:mfp}b).

\section{Summary and conclusions}\label{sec:summary}

In this work we have presented the simulation results of the
ESP event that was observed near Earth on 14 July 2012. 
This was done by using the MHD model EUHFORIA together with the energetic particle model PARADISE. EUHFORIA was used to model the background solar wind and the propagation of the CME through interplanetary space. 
The CME was modelled using the spheromak model of EUHFORIA, for which the magnetic and kinematic parameters had previously been derived by \citet{scolini19}. 
The resulting EUHFORIA simulation manages to capture both the shock arrival time and the solar wind speed jump at the shock passage  reasonably well, and a qualitative match is obtained between the observed and simulated flux rope passage  at Earth.
One important discrepancy is that in the simulation Earth connects magnetically to the shock wave at ${\sim} 0.5$ au, whereas the SEP data suggest that Earth established magnetic connection with the shock when the CME was still in the low corona. 
This difference is, at least partly, the result of the  IMF being a sub-Parker field due to the passage of an HSS on 11 July. 
Even though this HSS was captured by EUHFORIA, a sub-Parker IMF configuration is not reproduced by the model since EUHFORIA does not take into account effects of  differential rotation and magnetic footpoint motion, which are believed to produce sub-Parker spirals. 
Due to this late connection, the simulation cannot capture the onset phase of the SEP event.
This illustrates that one should be careful when using an MHD model such as EUHFORIA or a Parker spiral IMF to determine the magnetic connectivity of an observer.

A good agreement was obtained between the simulated and observed ESP event, especially at low energies ($<700$~keV). 
In our simulations, the energetic protons are accelerated at the CME-driven shock wave, assuming a seed population of 50 keV protons. 
The assumed seed population (see Eq.~\eqref{eq:j_inj}) may originate from the solar wind
suprathermal tail, especially near the shock wave where the solar wind gets shock-heated, hence producing a larger population of suprathermal protons \citep{prinsloo19}.
The good match between the observed and modelled intensity-time profiles confirms that the low-energy component of ESP events observed at 1~au may indeed be produced by the interplanetary acceleration of particles at the CME-driven shock wave. 
The simulations do not reproduce the intensity-time profiles for energies above ${\sim}1$~MeV. 
This energy threshold is close to the rollover energy of the observed energy spectrum. 
This correspondence suggests that the ESP component above 1 MeV depends on particle acceleration happening in the solar corona.
This will generate an extra, more energetic seed population that can be re-accelerated in the interplanetary medium along with the suprathermal ions of the solar wind.
In future work, we plan to extend our model to the corona to investigate this possibility further. 

In our simulation, the particle intensity-time profiles show a drop when the flux rope arrives. 
A similar sharp drop is observed in the EPAM data and indicates that the particles have limited access into the flux rope. 
In addition, the interaction between the flux rope and the IMF produces a secondary intensity peak, which is present both in the simulation and the data. 
The observed particle intensities do not drop to the pre-event background levels, indicating that some energetic particles reside inside the flux rope. 
Although not shown here, we tried to reproduce this by increasing the perpendicular mean free path in the simulation to allow the simulated SEPs to penetrate the flux rope more easily through cross-field diffusion. 
This, however, quickly diffuses the sharp drop in intensity that is seen at the beginning of the magnetic cloud. 
Hence, we believe that the observed SEPs inside the flux rope are likely not the result of a spatial diffusion process happening everywhere along the interface between the sheath and the flux rope. 
Instead, particles may enter the flux rope at locations where the sheath magnetic field reconnects with the flux rope magnetic field.  
This is in line with the findings of \citet{laitinen21}, who showed by using full orbit simulations that energetic particles penetrate flux ropes most easily near magnetic x points.

To reproduce the low-energy component of the ESP event, it was necessary to include a turbulent foreshock and sheath region in PARADISE, which we prescribed by using a parallel mean free path that decreases towards the shock wave.  
The parallel mean free path inside the foreshock was assumed to scale  with $1 / \cos^2\theta$ to take into account the fact that the turbulence upstream of a shock wave is expected to depend on the shock angle, as is the case for Earth's bow shock \citep[e.g.][]{eastwood05}.
The width of the foreshock region and the parallel mean free path at the shock surface were estimated from the observed ESP event at Earth and did not contain any explicit time dependence or dependence on the intensity of the modelled energetic particles. 

In order to transform a model such as EUHFORIA$+$PARADISE into a fully physics-based forecasting model for ESP or SEP events, it is important to have a model for the foreshock region and particle injection that does not rely on in situ observations of the ESP event. 
To achieve this, we plan to couple our model to the SOLPACS code \citep{afanasiev15}, which models particle acceleration at oblique shock waves, thereby taking into account the generation of Alfvén waves in the upstream region by the energetic particles themselves. 
Hence, this model provides a physics-based calculation of the parallel mean free path and its rigidity dependence inside the foreshock. 
This is imperative as the results presented in this work illustrate the sensitivity of the simulations on the assumed scattering conditions in the foreshock.


Solar energetic particle events are an important component of space weather since they can pose significant hazards for the health of astronauts and the microelectronics on board spacecraft. 
As a result, a profound understanding of the physical processes behind SEP events is necessary to be able to develop a reliable SEP forecasting tool. 
In this study, we presented the first attempt at using the EUHFORIA$+$PARADISE framework to reproduce an observed ESP event. 
With this work, we contribute to the ongoing worldwide effort \citep[e.g.][]{young21,li21} that aims at the development of a physics-based SEP forecasting tool. In the future, we plan to extend our study by modelling several of the recently observed multi-spacecraft SEP events of solar cycle 25.

\begin{acknowledgements}
We acknowledge the use of ACE data from the ACE Science Center and of SOHO/ERNE data from sepem.eu and of the SEPEM Reference Data Set version 2.00, European Space Agency (2016).
N.W.\ acknowledges funding from the Research Foundation -- Flanders (FWO -- Vlaanderen, fellowship no.\ 1184319N). A.A. and B.S. acknowledge the support by the Spanish Ministerio de Ciencia e Innovaci\'{o}n (MICINN) under grant PID2019-105510GB-C31 and through the ``Centre of Excellence Mar\'{i}a de Maeztu 2020-2023'' award to the ICCUB (CEX2019-000918-M).
C.S.\ acknowledges the NASA Living With a Star Jack Eddy Postdoctoral Fellowship Program, administered by UCAR's Cooperative Programs for the Advancement of Earth System Science (CPAESS) under award no.\ NNX16AK22G. D.L.\ acknowledges support from NASA Living With a Star (LWS) programs NNH17ZDA001N-LWS
and NNH19ZDA001N-LWS, and the Goddard Space Flight Center Heliophysics Innovation Fund (HIF) program.
The work in the University of Turku was performed under the umbrella of Finnish Centre of Excellence in Research of Sustainable Space.
This project has received funding from the European Union’s Horizon 2020 research
and innovation programs under grant agreement No.\ 870405 (EUHFORIA 2.0). These results were also obtained in the framework of the ESA project ``Heliospheric modelling techniques'' (Contract No.\ 4000133080/20/NL/CRS).
Computational resources and services used in this work were provided by the VSC (Flemish Supercomputer Centre), funded by the FWO and the Flemish Government-Department EWI. 
\end{acknowledgements}
\bibliographystyle{aa}
\bibliography{allpapers}

\begin{appendix}
\section{The shock finder algorithm}\label{app:shock_finder}
\begin{figure*}
    \centering
    \includegraphics[width=0.95\textwidth]{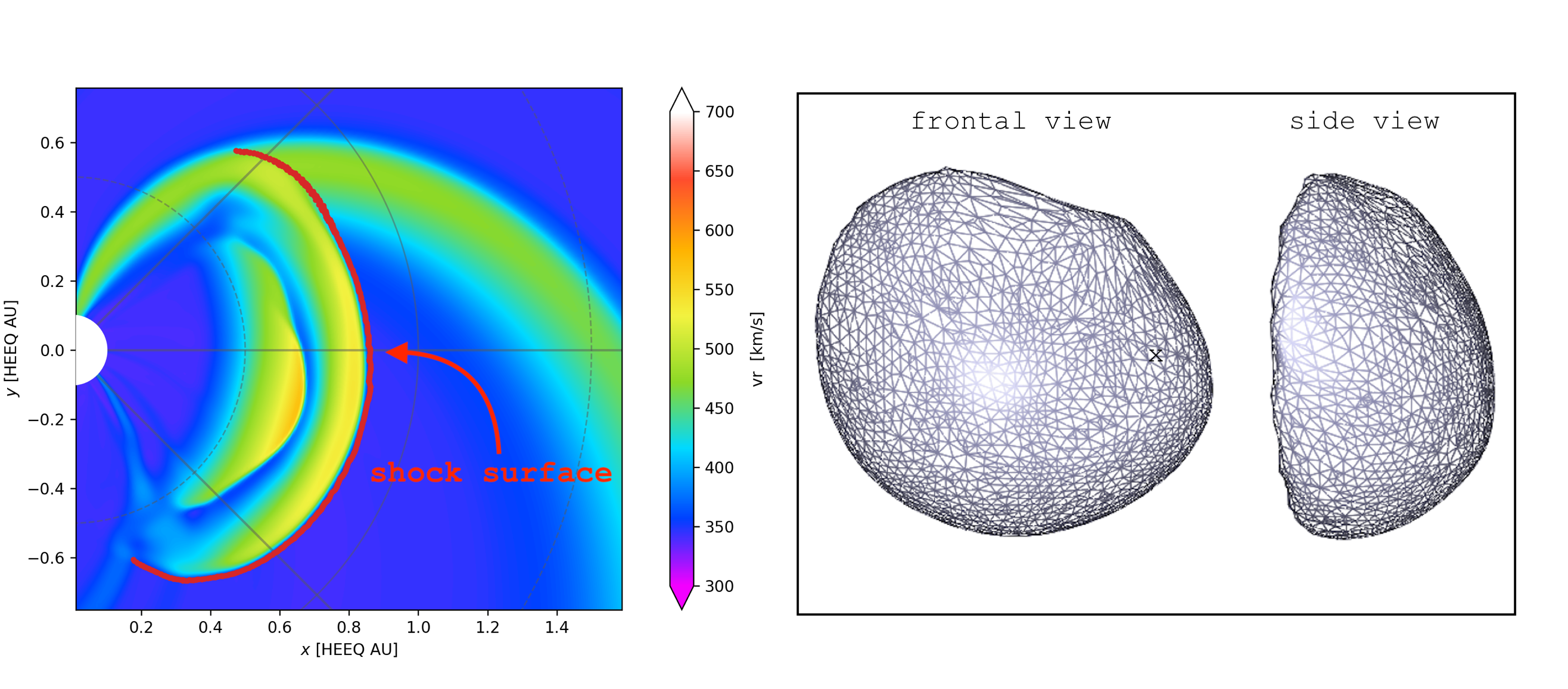}
    \caption{Illustration of the shock finder algorithm. \textit{Left panel:} Snapshot of a EUHFORIA simulation containing a cone CME. Shown is the radial speed in the solar equatorial plane. The red line indicates the shock surface, as determined by the MCA. \textit{Right panel:} 3D view of the shock surface computed by the MCA. The resolution has been reduced to clearly illustrate the triangle mesh.   }
    \label{fig:mc_illust}
\end{figure*}

In order to study CME-driven shocks in EUHFORIA simulations, we updated the shock finder tool that was developed by \cite{pomoell15} for a 2D MHD model. 
This new tool is used in PARADISE to both inject particles continuously at the shock front and prescribe the regions of increased particle scattering conditions in the foreshock and the sheath. 

The detection of a CME-driven shock wave is complicated by, for example, the presence of shocks associated with CIRs.
The shock-detection algorithm removes shocks driven by high-speed solar wind streams by subtracting the unperturbed background solar wind configuration from the solar wind in which the CME is propagating:
\begin{equation}
    q_{\rm rel}(t,r,\theta,\phi) \equiv \frac{q(t, r,\theta,\phi) - q_{\rm sw}(t,r,\theta,\phi)}{q_{\rm sw}(t,r,\theta,\phi)},
\end{equation}
where the subscript `sw' refers to a solar wind simulation that does not contain a CME and $q$ can be any plasma variable. 
With this definition, a CME corresponds thus to a $q_{\rm rel} \neq 0$ region in the simulation domain. 
Moreover, since the EUHFORIA solar wind  is steady state in a frame corotating with the sun, we have that  $q_{\rm sw}(t,r,\theta,\phi) = q_{\rm sw}(0,r,\theta,\phi-\Omega t)$, with $\Omega$ the solar rotation rate. 
Hence, $q_{\rm sw}(t)$ can be derived from a single snapshot, obtained, for example, just before the CME insertion. 

In a next step, we search for a specific point on the shock surface. 
If we are interested in deriving the entire shock surface,  $ q_{\rm rel}$ is calculated first along a radial line passing through the centre of the CME insertion location. 
If instead we are only interested in the shock properties at the cobpoint of a particular observer, $q_{\rm rel}$ is calculated along the IMF line crossing the observer. 
For $q$, we typically choose both the solar wind speed $V$ and the entropy $S \equiv P/\rho^\gamma$, where $P$ is the thermal pressure, $\rho$ the density, and $\gamma$ the adiabatic index. 
Following either a radial line segment or an IMF line from the outer boundary of the domain inwards, the shock point candidate $\vec{r}_1$ is determined as the first point for which both $V_{\rm rel}> \epsilon_V$ and $ S_{\rm rel} > \epsilon_S$, where $\epsilon_V$ and $\epsilon_S$  are predefined positive thresholds. 
Such non-zero thresholds are necessary due to the finite numerical accuracy of grid-based simulations.   
To ensure that $\vec{r}_1$ is indeed located at a compression wave, 
the algorithm checks whether the divergence of the solar wind velocity $\nabla\cdot\vec{V}$ is negative. 
This condition implies the presence of converging solar wind flows, for which shocks wave are extreme examples.

Next, the shock surface $S_1$ is determined around $\vec{r_1}$ by  computing it as the isosurface of $q_{\text{rel}}= \epsilon_{q}$, with $\epsilon_{q}$ a positive constant. 
As before, different choices for $q$ are possible. 
In this work, we choose $q_{\text{rel}} = -(\nabla\cdot \vec{V})_{\text{rel}}$, because the resulting isosurface encapsulates the simulated MHD shock wave. 
The isosurface is computed using the marching cubes algorithm (MCA), which generates the isosurface on a triangle mesh \citep{lewiner12}. 
Figure~\ref{fig:mc_illust} illustrates an application of the MCA to determine the shock surface generated by a cone CME in a EUHFORIA test simulation.
From the mesh generated by the MCA, it is straightforward to compute the shock-normal at any point on the shock surface. 
The shock angle $\theta_{B_n}$ can then  be determined by interpolating the upstream magnetic field to the shock surface.   
The shock speed $V_{\text{sh}}$ is calculated by determining the shock surface $S_2$ at a time instance that is typically $\Delta t = t_2-t_1 \sim 30$ minutes back in time. The shock speed at a point $\vec{r}_i\in S_1$ is then estimated as $\Delta x/\Delta t$, where $\Delta x$ is the minimal distance from point $\vec{r}_i$ to $S_2$.
The shock speed and the solar wind conditions just upstream of $\vec{r}_i$ are then used to solve the Rankine-Hugoniot equations, which give us the downstream solar wind conditions everywhere along the shock surface.

\section{Particle acceleration in PARADISE}\label{app:particle_acceleration}
\begin{figure*}
    \centering
    \includegraphics[width=0.9\textwidth]{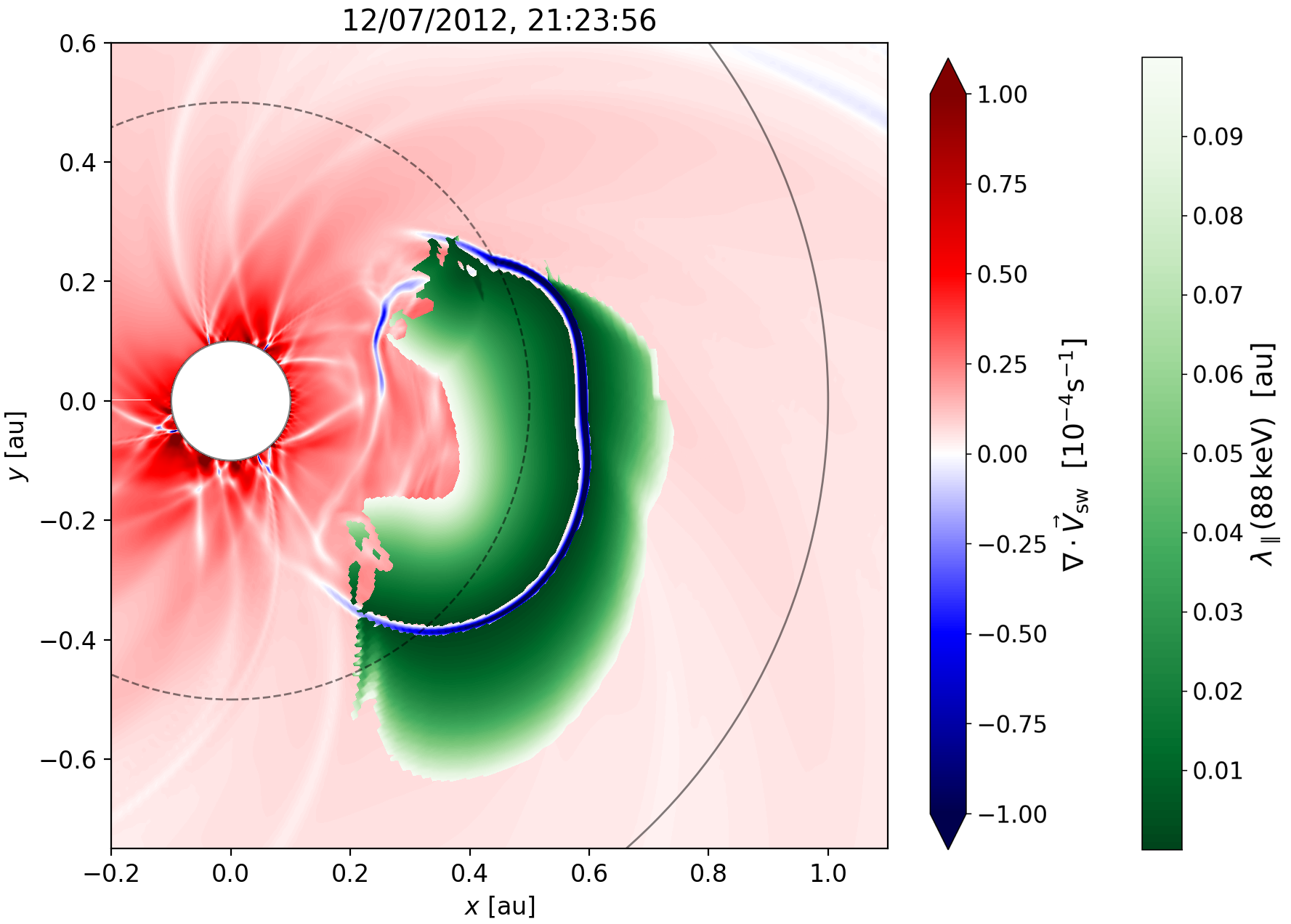}
    \caption{Snapshot of the parallel mean free path of 88 keV protons (green shades) and $\nabla\cdot \vec{V}_{\rm sw}$ (red and blue) at constant HEEQ latitude $\theta = 4^\circ$. The parallel mean free path is only shown where $\lambda_\parallel <  0.1$~au.  }
    \label{fig:mfp_on_sw}
\end{figure*}
PARADISE solves the FTE, which contains the necessary physics to describe particle acceleration in converging, accelerating, and shear flows \citep[e.g.][]{leRoux07,leRoux09}. 
Moreover, as illustrated by, for example, \cite{leRoux07}, the momentum term of the FTE contains the energy changes due to the particle drifts along the motional electric field. 
In this work, the FTE was solved without including a diffusion coefficient $D_{pp}$, where $p$ denotes the particle's momentum magnitude. This means that our simulations do not take into account any stochastic acceleration mechanism that may occur downstream of the shock wave \citep[e.g.][]{afanasiev14}.

When using a 3D MHD model such as EUHFORIA to simulate the evolution of a CME in the inner heliosphere, the shock wave generated in front of the CME will typically be much wider than a real interplanetary shock wave, due to the limited resolution of the numerical grid.
In such MHD simulations, the shock wave is thus  similar to a large amplitude compression wave, which can be characterised by the length scale \citep[e.g.][]{vainio18},
\begin{equation}\label{ch_cme:eq:Lc}
L_C \sim U_n / |\nabla \cdot \vec{U}|.
\end{equation}
Here, $\vec{U}$ denotes the velocity of the solar wind plasma measured in a reference frame co-moving with the shock wave, and the subscript `n' refers to the vector component measured along the shock normal.
The efficiency of the particle acceleration near compression waves  is  related to the particle diffusion length, which can be defined as \citep[e.g.][]{giacalone02}
\begin{equation}\label{ch_cme:eq:Ld}
L_D = \frac{\kappa_{nn}}{U_n} = \frac{v\lambda_n}{3U_n},
\end{equation}
where $\kappa_{nn}$ and $\lambda_n$ denote the particles' spatial diffusion coefficient and the  mean free path normal to the shock front, respectively. 
The latter is related to the particles' parallel and perpendicular mean free paths through 
\begin{equation}\label{ch_cme:eq:mfp_n}
\lambda_n = \lambda_\parallel\cos^2\theta_{Bn} + \lambda_\perp \sin^2\theta_{Bn}.
\end{equation}
If $L_D \gg L_C$, the particles can cross the compression wave multiple times and  gain energy through first-order Fermi acceleration \citep[e.g.][]{giacalone02,jokipii03,jokipii07}. 
For this acceleration process to happen, the diffusive length scale  upstream and downstream of the compression wave cannot be too large either (e.g. compared to the focusing length $L_F = \left(\nabla\cdot\vec{b}\right)^{-1}$), since otherwise particles will escape into the interplanetary medium \citep{battarbee11,vainio14}.
In the other limiting case, when  $L_D \ll L_C$, the particles will be closely tied to the scattering centres embedded in the solar wind and hence they are advected with the plasma flow through the compression wave\footnote{As seen from a frame co-moving with the shock or compression wave.}. As a result,  the particles will gain energy in an adiabatic fashion \citep[e.g.][]{giacalone05,vainio18,schwadron20}. 

In this work, we propagate the particles scatter free across the shock wave, which is identified as the $\nabla\cdot \vec{V}_{\rm sw} < 0 $ region in front of the CME (see Appendix~\ref{app:shock_finder}). 
This ensures that $L_D \gg L_C$ , since $L_{D} = +\infty$ across the CME-driven shock wave. 
The foreshock and sheath regions, where a small mean free path is prescribed according to Eqs.~\eqref{eq:mfp_IP} -- \eqref{eq:mfp_ss}, begin only outside the $\nabla\cdot \vec{V}_{\rm sw} < 0 $ region. 
Figure~\ref{fig:mfp_on_sw} shows the parallel mean free path for 88~keV protons together with $\nabla\cdot \vec{V}_{\rm sw}$. 
For clarity, we only show the parallel mean free path where $\lambda_\parallel < \lambda_\parallel^{\rm IP} = 0.1$~au. The simulated shock is visible as the blue arc inside the green shaded region.
The time step of the particles inside the modelled shock is calculated \textit{as if} $\lambda_\parallel = \lambda_\parallel^{\text{foreshock}}(d=0)$. 
This is done to avoid a large time step, which would transport the particles artificially far upstream or downstream of the shock in a single time step.
Finally, we refer to chapter~8 of \citet{wijsenPHD20} for a discussion of different PARADISE simulations in which the parallel mean free path is kept small across the CME-driven shock wave, hence leading to compressional acceleration. 

\end{appendix}

\end{document}